\documentclass[%
 aip,
 amsmath,amssymb,
 reprint,%
]{revtex4-1}

\usepackage{graphicx}
\usepackage{dcolumn}
\usepackage{bm}

\usepackage[utf8]{inputenc}
\usepackage[T1]{fontenc}
\usepackage{mathptmx}
\usepackage{etoolbox}
\usepackage{amsmath}

\makeatletter
\def\@email#1#2{%
 \endgroup
 \patchcmd{\titleblock@produce}
  {\frontmatter@RRAPformat}
  {\frontmatter@RRAPformat{\produce@RRAP{*#1\href{mailto:#2}{#2}}}\frontmatter@RRAPformat}
  {}{}
}%
\makeatother
\begin{document}

\preprint{AIP/123-QED}

\title[Suppression of laser beam’s polarization and intensity fluctuation via a
Mach-Zehnder interferometer with proper feedback]{Suppression of laser beam’s polarization and intensity fluctuation via a
Mach-Zehnder interferometer with proper feedback}
\author{Xiaokai Hou}
\affiliation{ 
State Key Laboratory of Quantum Optics and Quantum Optics Devices, and Institute of Opto-Electronics, Shanxi University, Tai Yuan 030006, Shanxi
Province, China
}%
  
\author{Shuo Liu}%

 \altaffiliation{Present Address: Key Laboratory of Laser $\&$ Infrared System of
Ministry of Education, Shandong University, Qing Dao 266000, Shandong
Province, China.}
\affiliation{ 
State Key Laboratory of Quantum Optics and Quantum Optics Devices, and Institute of Opto-Electronics, Shanxi University, Tai Yuan 030006, Shanxi
Province, China
}%

\author{Xin Wang}

\affiliation{ 
State Key Laboratory of Quantum Optics and Quantum Optics Devices, and Institute of Opto-Electronics, Shanxi University, Tai Yuan 030006, Shanxi
Province, China
}%

\author{Feifei Lu}
 
\affiliation{ 
State Key Laboratory of Quantum Optics and Quantum Optics Devices, and Institute of Opto-Electronics, Shanxi University, Tai Yuan 030006, Shanxi
Province, China
}%

\author{Jun He}
 
\affiliation{ 
State Key Laboratory of Quantum Optics and Quantum Optics Devices, and Institute of Opto-Electronics, Shanxi University, Tai Yuan 030006, Shanxi
Province, China
}%
\affiliation{ 
Collaborative Innovation Center of Extreme Optics, Shanxi University, Tai Yuan 030006, Shanxi Province, China
}%

\author{Junmin Wang}

 \altaffiliation{corresponding author. E-mail: wwjjmm@sxu.edu.cn ORCID : 0000-0001-8055-000X}
 
\affiliation{ 
State Key Laboratory of Quantum Optics and Quantum Optics Devices, and Institute of Opto-Electronics, Shanxi University, Tai Yuan 030006, Shanxi
Province, China
}%
\affiliation{ 
Collaborative Innovation Center of Extreme Optics, Shanxi University, Tai Yuan 030006, Shanxi Province, China
}%
\date{\today}

\begin{abstract}
Long ground-Rydberg coherence lifetime is interesting for implementing high-fidelity quantum logic
gates, many-body physics, and other quantum information protocols. But, the potential well formed
by a conventional far-off-resonance red-detuned optical-dipole trap that is attractive for ground-state
cold atoms is usually repulsive for Rydberg atoms, which will result in the rapid loss of atoms
and low repetition rate of the experimental sequence. Moreover, the coherence time will be sharply shortened due to the residual thermal motion of cold atoms. These issues can be addressed by an one-dimensional magic lattice trap and it can form a deeper potential trap than the traveling wave optical dipole trap when the output power is limited. And these common techniques for atomic confinement generally have certain requirement on the
polarization and intensity stability of the laser. Here, we demonstrated a method to suppress both the
polarization drift and power fluctuation only based on the phase management of the Mach-Zehnder
interferometer for one-dimensional magic lattice trap. With the combination of three wave plates and
the interferometer, we used the instrument to collect data in the time domain, analyzed the fluctuation
of laser intensity, and calculated the noise power spectral density. We found that the total intensity
fluctuation composed of laser power fluctuation and polarization drift was significantly suppressed,
and the noise power spectral density after closed-loop locking with typical bandwidth 1-3000 Hz was
significantly lower than that under the free running of the laser system. Typically, at 1000 Hz, the
noise power spectral density after locking was about 10 dB lower than that when A Master Oscillator Power Amplifier (MOPA) system free running.
The intensity-polarization control technique provides potential applications for atomic confinement
protocols that demand for fixed polarization and intensity 
\end{abstract}

\maketitle

\section{\label{sec:level1}Introduction}

For various atomic manipulation experiments, such as single photon source$^{1-5}$, quantum dynamics
based on Rydberg states $^{6-10}$ and electric field detection based on atoms $^{11-13}$, strong confinement optical dipole trap (ODT) of atoms is usually employed. In these applications,high power laser with fixed polarization and relatively stabled intensity normally is used to confine atoms. Common experimental setup for the laser power stabilization were based on the active feedback loop which used acousto-optic modulator (AOM) $^{14-17}$ or electro-optic modulator (EOM) $^{18}$ as the actuator. In 2020, AOM and EOM were combined to broaden the bandwidth of laser intensity noise stabilization to 1MHz by Ni et. al.$^{19}$. At present, the feedback loop based on AOM has some disadvantages. For example, bragg diffraction of AOM will seriously affect the spot quality of first-order diffraction light, and the power utilization of the system will be limited by the diffraction efficiency of AOM. The common electro-optic intensity modulator (EOIM) with input and output tailed fiber is efficient, but not suitable for high-power
applications. Moreover, the schemes mentioned above can observably suppress the power fluctuation of laser beam, but the reduction for the drift of laser’s polarization is still not be effectively achieved . Here we demonstrate an experimental scheme based on the Mach-Zehnder interferometer (MZI) for actively suppress both the fluctuation of power and polarization of laser beam. By properly manipulated
phase difference between two paths, the output fraction of MZI account for the majority laser power while its intensity fluctuation in the time domain has been reduced dozens of times compared with the free-running case, and the noise power spectral density (NPSD) has been decreased in the range of 1-3000 Hz in the frequency domain. Such a stable
system can certainly meet the needs of various applications, such as experiments where the lifetime of cold atoms is highly desirable.

\section{\label{sec:level2}Theoretical background}
\subsubsection{\label{sec:level3}Magic optical dipole trap for cesium $6S_{1/2}$ ground state and $84P_{3/2}$ Rydberg state}

Recently, a new experimental scheme which used interferometer as the actuator of the feedback loop 
has been proposed $^{20}$. Considering the light intensity requirement of the ODT, the MZI can satisfy the power requirement of ODT without affecting spot quality of output light, therefore the experimental setups of constructing the blue-detuning optical trap reported by Yelin et. al $^{21}$ and Isenhower et. al $^{22}$ both concentrate on the MZI. The intensity of the output laser mainly depends on the phase difference between two arms of the MZI, therefore it can be used as a power stabilizer in some experiments $^{23}$. Due to the particularity of the output fraction of MZI, the combination of interferometer and polarizer can realize the fixed polarization, high proportion output and high intensity stability. It is obviously
useful for the experiment of optical trap. The potential of
ODT $U$ can be expressed as:
\begin{flalign}
U=-\frac{\alpha}{2 \epsilon_0 c} \frac{2 P}{\pi \omega_0^2}  
\end{flalign}

Where $\alpha$ is the induced polarizability of the target state, $\epsilon_0$ is the permittivity of vacuum, $c$ is the speed of light, $P$ is the intensity of laser, $\omega_0$ is the radius of the spot at the focal point after the laser is focused by a lens . As shown in the Eq. (1), if the power of the 1879 nm laser is fluctuant, the resulting trap depth will be changed. Thus, the lifetime of the trapped atom will be severely affected by the presence
of the heating mechanism$^{5,17,24}$.
\begin{figure}[h]
    \centering
    \includegraphics[width=0.4\textwidth,height=0.3\textwidth]{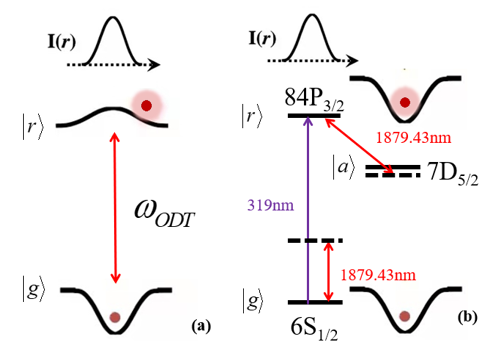}
    \caption{Diagram of the light shift induced by the ODT and MODT. The intensity of laser which is intensly focused is still Gaussian, and the closer to the center of beam, the stronger the intensity of laser. The resulting trap depth or light shift is spatially dependent (a) The ODT is attractive for ground states, but usually repulsive for highly-excited Rydberg states because almost all strong dipole transitions connected Rydberg state and the lower states have longer wavelength than that of ODT laser. (b) The direct single-photon excitation scheme from cesium $|g\rangle$=$|6S_{1/2}\rangle$  to  $|r\rangle$=$|84P_{3/2}\rangle$ coupled by a 319 nm ultraviolet laser. A 1879.43 nm laser is also tuned to the blue side of the $|r\rangle$ $\Longleftrightarrow$ $|a\rangle$=$|7D_{5/2}\rangle$ auxiliary transition to equalize the trapping potential depth of the $|g\rangle$ and $|r\rangle$ state, which is so called magic ODT (MODT).}
\end{figure}

In most of the experiments of cold atoms involving
confinement of ground-state atoms in an ODT and Rydberg
excitation, cold atomic sample is prepared in an ODT to hold them in a fixed position in a significantly long time. The potential formed by a conventional far off-resonance red-detuned ODT is attractive for the ground-state atoms, but usually repulsive for highly-excited Rydberg atoms, leading that Rydberg atoms normally cannot be confined in the conventional ODT (Fig. 1(a)). Therefore, in the follow-up experiments, we will face the following two problems: (1)
if switching off the ODT during Rydberg excitation and
coherent manipulation, it will result in atomic dephasing due to the thermal diffusion of the atoms and the extremely low repetition rate of the experimental sequence; (2) if the ODT
remains operation, it may cause a low Rydberg excitation efficiency of atoms as the transition frequency is spatially position-dependent on the excitation laser. The solution is to find an ODT such that the ground-state atoms and the desired highly-excited Rydberg atoms can experience the same potential, that is, the potential generated by the ODT
is a potential well for both the ground-state atoms and the desired highly-excited Rydberg atoms, and is attractive to atoms in both states. So, the above-mentioned aspects (1) and (2) can be solved. In Fig.1(b), the direct single-photon excitation scheme from cesium $|g\rangle$=$|6S_{1/2}\rangle$  to  $|r\rangle$=$|84P_{3/2}\rangle$  coupled by a 319 nm ultraviolet laser. A 1879.43 nm laser is also tuned to the blue side of the $|r\rangle$ $\Longleftrightarrow$ $|a\rangle$=$|7D_{5/2}\rangle$   auxiliary transition to equalize the trapping potential depth of the $|g\rangle$ and $|r\rangle$ state. The specific calculation process is not described here. For details, please refer to the reference $^{25,26}$.

\subsubsection{\label{sec:level3}Theoretical analysis of MZI}

It is obvious that the MODT is not enough to meet
the need of extremely long coherence time in subsequent
experiments. The cold atoms trapped in the MODT still have residual thermal motion, which causes violent collisions that heat the atoms and cause them to escape from the trap. We will further construct one-dimensional magic lattice trap (1D-MLT), and combine the advantages of lattice and magic conditions, so as to prolong the coherence time of the ground-Rydberg state of cold atoms. Of course, the 1D-MLT also needs to suppress its power fluctuation. Because the power of the laser used in the 1D-MLT fluctuates in the time domain, will directly shorten the coherence
lifetime of the cold atom. Therefore, we use the MZI to
suppress the power fluctuation.

As shown in Fig. 2 (a), $I_{out1}$ and $I_{out2}$ are the intensity
of two output paths of the interferometer respectively; $R_1$,
$T_1$, $R_2$, $T_2$ are the reflectivity and transmittance of input and output beam splitters plate respectively. The two output channels of the interferometer can be expressed as Eq. (2)
and (3):
\begin{flalign}
 I_{out1}&=R_1^2 R_2^2+T_1^2 T_2^2+2 R_1 R_2 T_1 T_2 cos(\frac{2 \Delta L}{\lambda}+\pi)\\   
 I_{out2}&=R_1^2 R_2^2+T_1^2 T_2^2+2 R_1 R_2 T_1 T_2 cos(\frac{2 \Delta L}{\lambda})   
\end{flalign}
Therefore, the laser intensity output of the interferometer can be controlled by adjusting the driving voltage of PZT due to the correlation between the output transmittance $I$ and optical path difference $\Delta L$. In Fig. 2 (b), the interference
fringes generated by splitters with different splitter ratio is simulated and analyzed by Mathematica. The splitter ratio shown by the first line of Fig. 2(b) is 90/10, the second is 70/30, the third is 60/40, and the last is 50/50.
\begin{figure}[h]
    \centering
    \includegraphics[width=0.48\textwidth,height=0.33\textwidth]{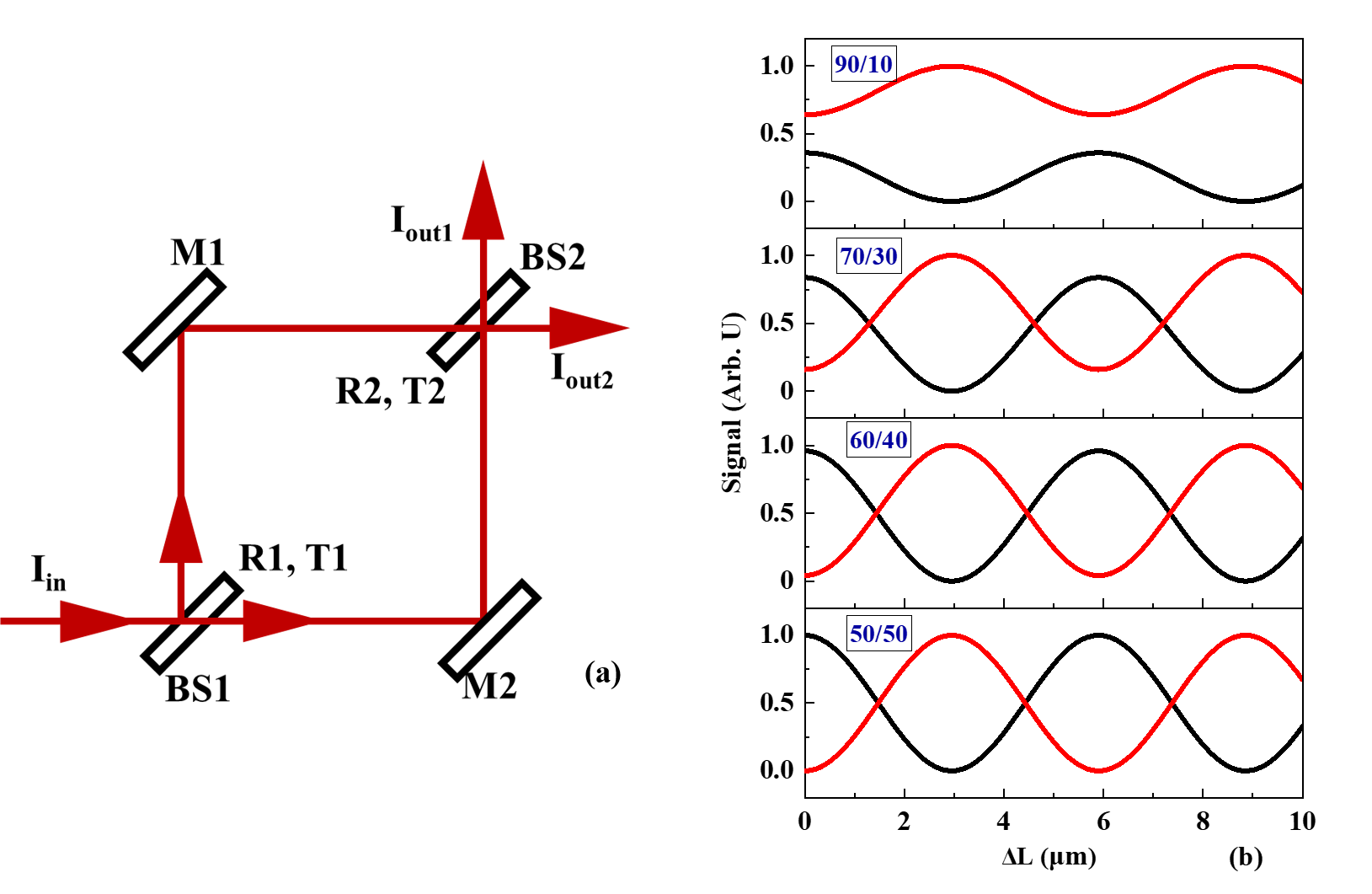}
    \caption{Diagram of MZI and interference fringes of two channels of the MZI are simulated and analyzed theoretically. (a) MZI consists of two beam splitter plates (BS1 and BS2) and two high-reflectivity mirrors (M1 and M2). $I_{in}$ is the intensity of the incident light field, $I_{out1}$ and $I_{out2}$ are the intensity of the outgoing light field at BS2. (b) Normalized signal as a function of difference of optical path $\Delta L$ for different splitter ratio.This ratio is both $R_1/T_1$ and $R_2/T_2$, because the BS1 and BS2 that are used in the MZI are same. The solid red and black lines represent the interference fringes of the two output channels of MZI, respectively.}
    \label{fig:my_label}
\end{figure}

\section{Experimental setup}
The laser intensity stabilization setup is shown in Fig. 3. A MOPA system consists of a 1879-nm butterfly packaged laser diode and a Thulium Doped Fiber Amplifier (TmDFA) which has maximum output $\sim$ 3 W. With a free space polarization controller based on three waveplates ( $\lambda / 4$, $\lambda / 2$ and $\lambda / 4$ ), polarization fluctuation of 1879 nm beam is suppressed initially.
 The laser is injected into a MZI which is constructed by a 50/50 beam splitter plate (BS1) that divides the incident light into two beams with equal intensity and the different phase, a high-reflectivity mirror (M1) that reflects one beam, a mirror (M2) attached to a PZT that emits the other, and a beam splitter plate (BS2) that the two beams are finally combined. The interferometer has two output channels and each channel can be used for dynamic feedback to make
the system more stable, and the output of this channel can then be used for subsequent experiments. The photodetector (PD1) is mounted behind a glass slice (GS1) of 1879 nm for sampling a little fraction of light for in-loop feedback. The DC voltage signal output by the PD1 is injected into Proportional Integral Differential (PID) amplifier after passing through a low-pass filter (LPF). The input signal of PID controller is subtracted from the PID Set Point, which is an artificially set reference DC voltage. The output signal of PID, that is the real-time difference between the detector signal and the reference DC voltage is added with the scanning signal (triangular wave) and amplified by the high voltage (HV) amplifier as the driving voltage of the PZT. The output power of interferometer can therefore be controlled by manipulating the driving voltage of PZT, and we expect that both power and polarization fluctuation for 1879 nm laser are suppressed. And another photodetector (PD2) is mounted in order to independently monitor the intensity stability of the output linear polarization laser. The output signal of PD2 is then injected into the Data Acquisition System (Keithley, DAQ-6510) in order to analyze and monitor the intensity fluctuation of the laser in the time domain and calculate the
NPSD based on the measured optical power fluctuation data. Undoubtedly, the little fraction of the far-infrared laser is reflected by glass slice (GS2) and received by the PD2 and the majority of laser is transmitted and focused in a cesium magneto-optical trap (Cs-MOT) for the construction of the ODT.
\begin{figure}[h]
    \centering    \includegraphics[width=0.5\textwidth,height=0.32\textwidth]{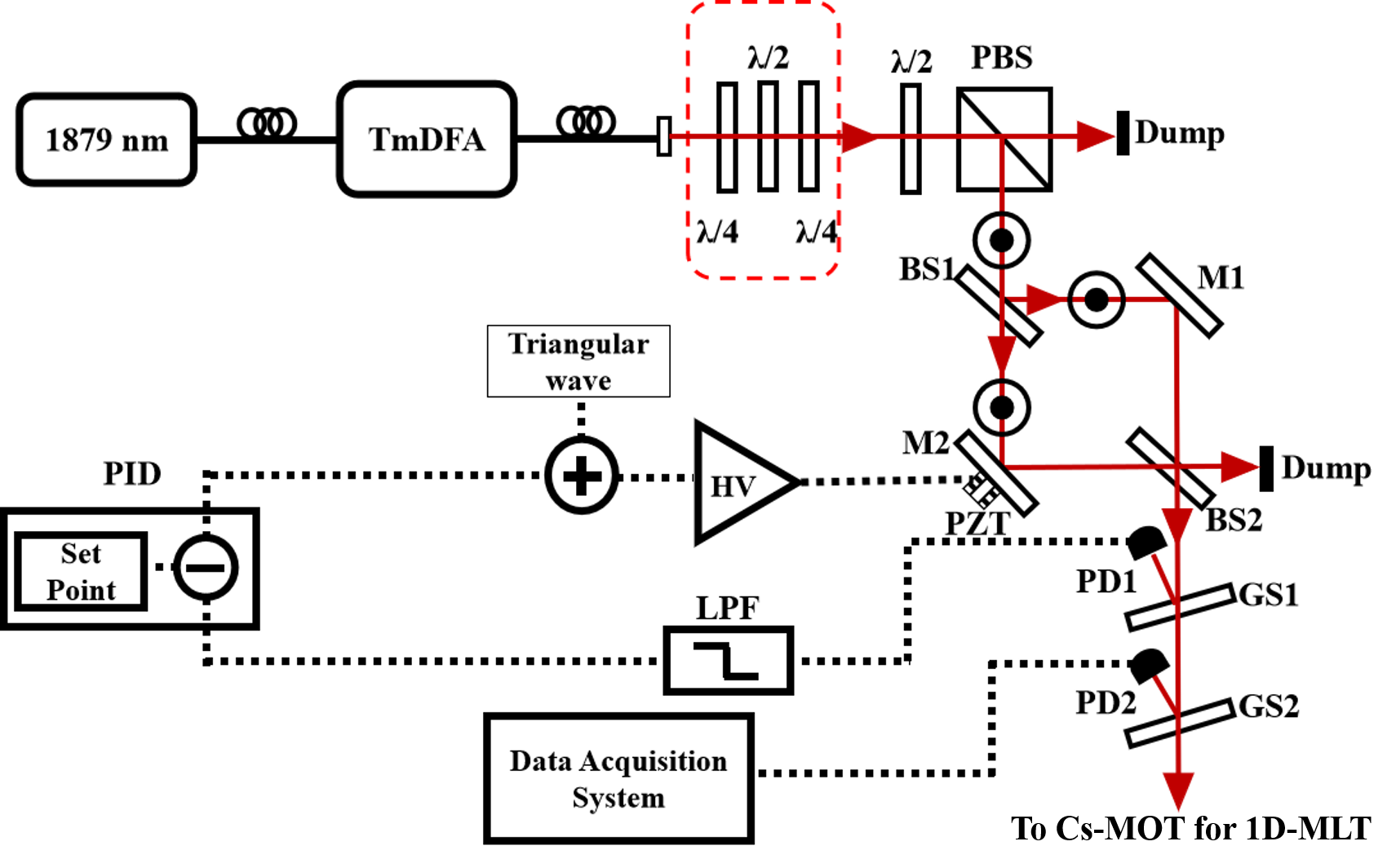}
    \caption{Experimental setup for intensity stabilization system. The dynamic stability of laser intensity of 1879 nm MOPA system is realized by MZI, and the fluctuation of laser intensity is monitored and analyzed in time domain and frequency domain.$\lambda / 2$: half-wave plate; $\lambda / 4$: quarter-wave plate; PBS: polarization beam splitting cube; BS: beam splitting plate; GS: glass slice; M1/M2: high-reflectivity mirror; PD: photodetector; LPF: low-pass filter; PID: Proportional Integral Differential amplifier; HVA: high voltage amplifier.}
    \label{fig:my_label}
\end{figure}
\section{Experimental results and discussion}

Fig. 4 shows, the interference fringes obtained by scanning triangular waves with 50/50 beam splitter ratio in the experiment, in which the interference contrast is 95$\%$. In theoretical simulation, an interference fringe with an interference contrast of 99.9$\%$ can be obtained by using a 50/50 beam splitter plate, but the best interference contrast is not achieved in experiment, probably due to the following two reasons: first, the spatial mode of the two lasers is not exactly same; second, the polarization of the two lasers may be slightly different.
\begin{figure}[h]
\centering
    \includegraphics[width=0.45\textwidth,height=0.35\textwidth]{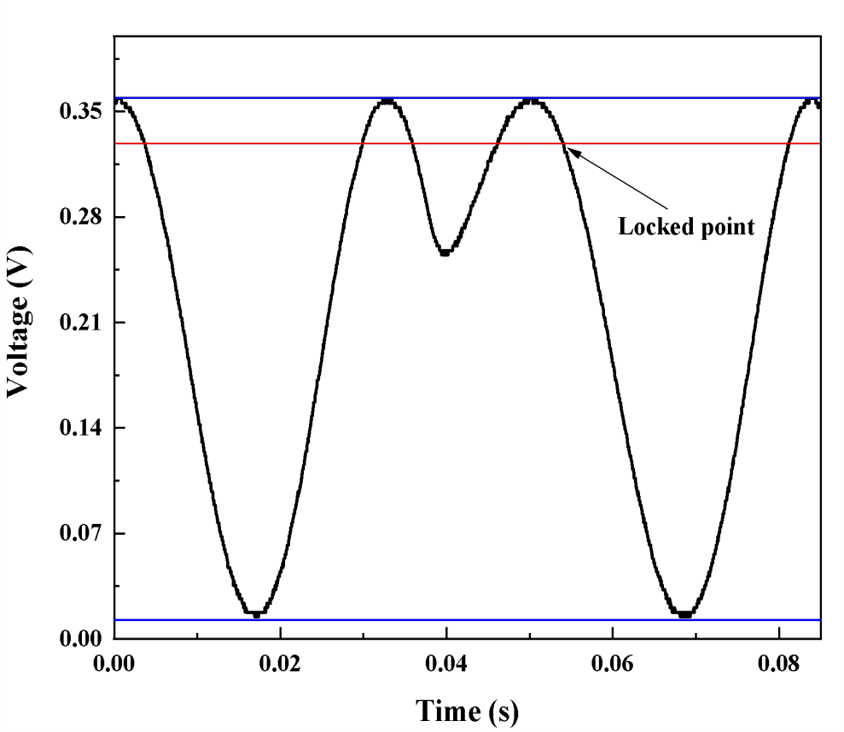}
\caption{Interference fringe of the MZI. In the experiment, a 50/50 beam splitter plate is used, and the PZT is driven by scanning triangular wave, so that the phase difference between the two arms is generated, then the interference fringes are generated.}
    \label{fig:my_label}
\end{figure}

Considering the requirement of constructing dipole trap with this laser source, the polarization of 1879 nm laser should be fixed, so PBS is usually inserted in the light path to fixed the polarization of light. Even though the scheme is effective, an inevitable defect exists in this scheme is that the polarization fluctuation of light will couple with the intensity fluctuation through this polarization element. As the measurement of which the intensity for 1879 nm laser after a PBS, although the power fluctuation of 1879 nm TmDFA itself is not obvious, the intensity fluctuation behind the PBS becomes obvious and the results is shown in Fig. 5(a). We monitor the laser intensity for about 30 minutes in the time domain, with a large fluctuation of about $\pm$14.2$\%$. The huge intensity fluctuation will significantly affect the power utilization of the stable system. To maximize the power utilization, three wave plates are used to suppressed the power fluctuation initially. After proper adjustment, measurement result of laser intensity fluctuation after PBS is shown in Fig. 5 (b). Fluctuation of laser polarization has been reduced significantly. 
\begin{figure*}
\centering
    \includegraphics[width=0.9\textwidth,height=0.35\textwidth]{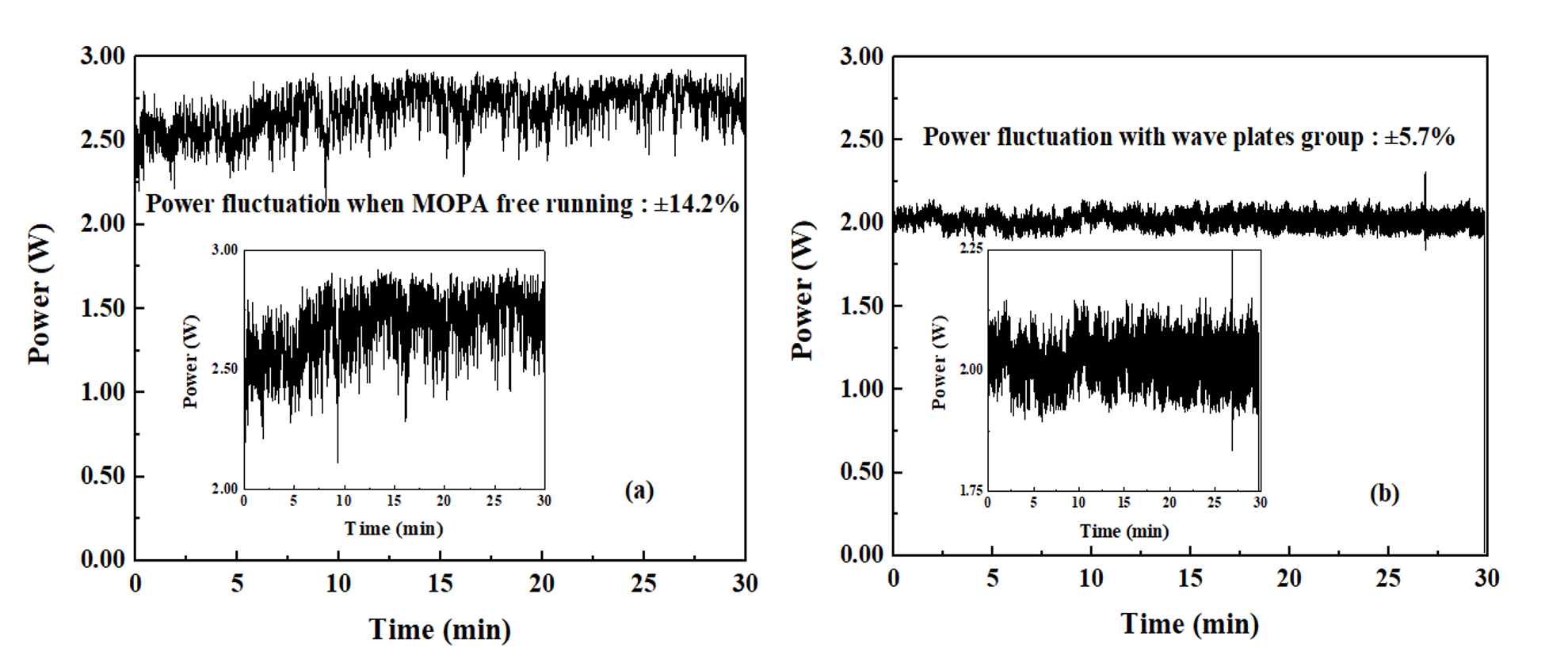}
\caption{(a) The power fluctuation of free running 1879 nm MOPA system . Through 30 mins of measurement, the power fluctuation is roughly $\pm$14.2$\%$. The inset is zoomed in on the vertical axis to 2.00$\sim$3.00 W, and shows the intensity fluctuation in 30 mins. (b) The power fluctuation after three wave plates. We thought that the polarization fluctuation is initially suppressed by these plates. Similarly, after 30 minutes of measurement, the intensity fluctuation is approximately $\pm$5.7$\%$. And, the vertical axis range of the inset becomes 1.75$\sim$2.25 W, the range of horizontal axis is still 0$\sim$30 mins.}
\end{figure*}
\begin{table*}
    \centering
    \begin{tabular}{c|c|c|c|c|c}
    \hline
     Category & $P_{ODT}$(mW) & $\Delta P$(mW) &  Gaussian radius after focused ($\mu$m) & $U_{dip}$($\mu$K) & $\Delta U_{dip}$($\mu $K) \\
     \hline
MOPA free running & 1500 & $\pm$213.0 ($\pm$14.2$\%$) & 20 & $-$1000 & $\pm$140 \\
With wave plate group & 1200 & $\pm$68.4 ($\pm$5.7$\%$) & 20 & $-$800 & $\pm$45 \\
After MZI is locked & 1100 & $\pm$3.3 ($\pm$0.3$\%$) & 20 & $-$700 & $\pm$2 \\
    \hline
    \end{tabular}
    \caption{The typical maximum trap depth and fluctuation of 1879.43nm 1D-MLT for cesium atoms under different power fluctuations are calculated.}
\end{table*}
Then the initial-stabled laser has been injected in the combine system of MZI and another PBS, here 
the transmittance of the interferometer is locked up to 90$\%$ in order to improve the power utilization. Then the intensity fluctuation probed by the out-of-loop detector PD2 is shown in Fig. 6. As shown below, the intensity fluctuation of output linear polarized laser is reduced to $\pm$0.3$\%$, that is much better than the fluctuation of direct TmDFA-PBS output. At this stability, both fluctuation of laser power and polarization will no longer have a significant influence on the parameter of dipole trap. 
\begin{figure}
    \centering
    \includegraphics[width=0.45\textwidth,height=0.35\textwidth]{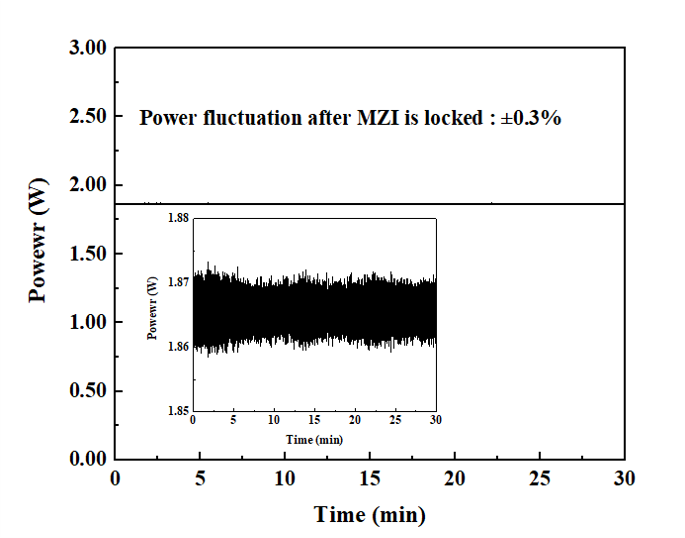}
    \caption{The intensity fluctuation of 1879 nm laser on the bright fringe of the MZI. By inter-of-loop locking, the phase difference between the two arms is dynamically compensated , and the power fluctuation is significantly suppressed , through 30 mins of measurement, the power fluctuation is roughly $\pm$0.3$\%$. The vertical axis of the inset has been enlarged with a range of 1.85$\sim$1.88 W, and shows the 30-min measurement.}
\end{figure}

As shown in Table 1, for the 1879 nm 1D-MLT, if the laser is focused through a lens to $\sim$ 20 $\mu$m. and the incident laser power at the cold atom is about 1.5 W, so the maximum depth of the 1D-MLT is $-$1000 $\mu$K and the typical trap depth fluctuation is $\pm$140 $\mu$K. When the laser power decreases after the initial suppression of the wave plate group or the closed-loop locking of the MZI, the corresponding typical trap depth is about $-$800 $\mu$K and $-$700 $\mu$K respectively. And the effective temperature of the cold atoms which are transferred from MOT to 1D-MLT will be slightly higher, about 100 $\mu$K, but the decrease of trap depth caused by the suppression of power fluctuation will not affect the capture of the cold atoms. However, the residual fluctuation of laser power still exists, which will lead to the typical trap depth fluctuation of $\pm$45 $\mu$K and $\pm$2 $\mu$K respectively.

The collected time-domain voltage signals are used to calculate the NPSD. As shown in Fig. 7, the horizontal range is determined by the sampling rate. In the experiment, we selected sampling rate of 10000 Hz according to the actual situation, so the horizontal axis in Fig. 7 ranges from 1 to 5000Hz. In addition, we believe that the feedback bandwidth of the system should be at the level of kilohertz due to the limitation of PZT in the MZI. Therefore, the sampling rate can fully meet the requirement of representing the feedback bandwidth of the system.

The NPSD after closed-loop locking from 1-3000Hz is significantly lower than that under the free running of the MOPA system. It can be proved that the MZI plays an obvious role in the power stability of the system. In order to further broaden the feedback bandwidth and improve the inhibitory effect,
we assume that the arm length of the MZI is $L$ and the angular frequency of the laser is $\omega_0$, then the distance of the laser going through the MZI is $L$ and the phase shift generated is$^{27}$
\begin{flalign}
  \Phi_0 (t)=\omega_0 t=\omega_0 \frac{L}{c}  
\end{flalign}

  $\Phi_0$ is a constant, and the magnitude is proportional to $L$ . When the PZT is scanned, we introduce  to characterize small changes in phase. For simplicity, we assume that a sine wave is used to scan the PZT, and the amplitude of the sine wave is $h_0$ and the angular frequency is $\omega_s$, so the sine wave can be expressed as 
  \begin{flalign}
  h(t)=h_0 cos(\omega_s t) 
  \end{flalign}
  
  So, the phase shift of the entire system can be written as
   \begin{flalign} 
   \nonumber 
\Phi &=\Phi_0 (t)+\delta \phi\\
 \nonumber
 &=\frac{\omega_0 L}{c}+\frac{\omega_0}{2}\int_{t-\frac{L}{c}}^{t} h_0 cos(\omega_s t) dt\\
 \nonumber
&=\frac{\omega_0 L}{c}+\frac{h_0}{2} \frac{\omega_0}{\omega_s}\left\{sin(\omega_s \frac{L}{c})-sin[\omega_s(t- \frac{L}{c})]\right\}\\
 &=\frac{\omega_0 L}{c}+ h_0 \frac{\omega_0}{\omega_s}sin(\omega_s \frac{L}{2 c}) cos[\frac{\omega_s}{2} (t-\frac{L}{c})]
\end{flalign}  

because $\frac{L}{2 c}\ll 1$
\begin{flalign}
 h_0 \frac{\omega_0}{\omega_s}sin(\omega_s \frac{L}{2 c})=\frac{h_0 \omega_0}{2} \frac{L}{c}
\end{flalign}
\begin{flalign}
\delta \phi \sim \frac{h_0 \omega_0}{2} \frac{L}{c}    
\end{flalign}

As shown in Eq. (8), if the arm length $L$ of the MZI is increased, $\delta \phi$ of the system can be increased. Thus, the detection sensitivity of the system can be improved and the detection effect of the MZI for phase can be better. Increasing the arm length of the MZI will cause extra noise due to the insufficient stability of the system. 

However, such noise can be solved through the isolation platform and system temperature control. We can add F-P cavity on the two arms of the MZI. F-P cavity can fold up the optical path, greatly increase the distance of light in the MZI, and do not need to occupy a large area.
\begin{figure}
    \centering
    \includegraphics[width=0.49\textwidth, height=0.38\textwidth]{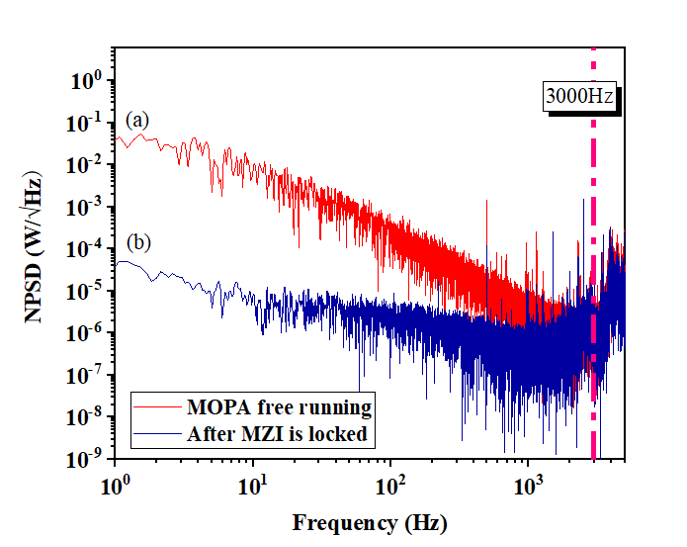}
    \caption{Intensity noise of 1879 nm laser as a function of analyze frequency. (a): The solid black line represents the NPSD when the 1879nm laser system is running freely without passing through the wave plate group. (b): The solid blue line represents the NPSD of the 1879nm laser system after closed-loop locking by the MZI.}
    \label{fig:my_label}
\end{figure}
\section{Conclusions}
In summary, we have demonstrated the reduction of the power and polarization fluctuation for 1879 nm laser based on the cooperation of three wave plates and a MZI. The intensity fluctuation $\sim$ $\pm$14.2$\%$ after the combination of MOPA system and PBS is reduced to $\sim$ $\pm$0.3$\%$ with locked MZI. And after MZI is locked, the NPSD is lower than that under free running in the range of 1-3000 Hz. Typically, at 1000 Hz, the NPSD after MZI is locked is about 10 dB lower than that when MOPA free running. The system can not only withstand high power injecting laser, but also can stabilize both power fluctuation and polarization fluctuation without affecting the quality of light beam for the low-loss output light. The laser power utilizing efficiency can be further improved by improving the transmittance of locked interferometer or improving the interference visibility.

It is expected that Rydberg atoms can have long coherence lifetime in subsequent experiments involving Rydberg dressed ground state. On one hand, we can use the 1879-nm MOPA system to implement a 1D-MLT, which can both eliminate the position-dependent light shift to capture Rydberg-state atoms in optical tweezer like the ground-state atoms and attenuate collisions between cold atoms caused by residual thermal motion to prolong the coherence time of the Rydberg atoms. On the other hand, we propose an upgraded interferometer, that is, adding a F-P cavity to each arm of the interferometer, and using the reflection of beam in the cavity, the arm length can be extended at least dozens of times, to improve the phase measurement sensitivity of the interferometer and improve the power stability. 

\section*{Funding}
This research was financially funded by the National Key R $\&$ D Program of China (2021YFA1402002), the National Natural Science Foundation of China (11974226, and 61875111).

\section*{References} 

$^1$ M. Endres, H. Bernien, A. Keesling, H. Levine, E. R. Anschuetz, A. Krajenbrink, C. Senko, V. Vuletic, M. Greiner, and M. D. Lukin. Atom-by-atom assembly of defect-free one-dimensional cold atom array[J]. Science, {\bf 354}, 1024, (2016).\\
$^2$ H. Kim, W. Lee, H.-G. Lee, H. Jo, Y. H. Song, and J. Ahn. In situ single-atom array synthesis using dynamic holographic optical tweezer[J]. Nature Commun., {\bf 7}, 1-8, (2016).\\
$^3$ B. Darquié, M. P. A. Jones, J. Dingjan, J. Beugnon, S. Bergamini, Y. Sortais,
G. Messin, A. Browaeys, and P. Grangier. Controlled single-photon emission from a single trapped two-level atom[J]. Science, {\bf 309}, 454-456, (2005).\\
$^4$ V. Leong, S. Kosen, B. Srivathsan, G. K. Gulati, A. Cere, and C. Kurtsie. Hong-ou-mandel interference between triggered and heralded single photons from separate atomic system[J]. Phys. Rev. A, {\bf 91}, 063829, (2015).\\
$^5$ B. Liu, G. Jin, J. He, and J. M. Wang. Suppression of single cesium atom heating in a microscopic optical dipole trap for demonstration of an 852nm triggered single-photon source[J]. Phys. Rev. A, {\bf 94}, 013409, (2016).\\
$^6$ Y. O. Dudin and A. Kuzmich. Strongly interacting Rydberg excitations of a cold atomic gas[J]. Science, {\bf336}, 887–889, (2012).\\
$^7$ Y.-Y. Jau, A. M. Hankin, T. Keating, I. H. Deutsch, and G. W. Biedermann. Entangling atomic spins with a rydberg-dressed spin-flip blockade[J]. Nature Phys., {\bf12}, 71–74, (2016).\\
$^8$ E. Urban, T. A. Johnson, T. Henage, L. Isenhower, D. D. Yavuz, T. G. Walker, and M. Saffman. Observation of Rydberg blockade between two atoms[J]. Nature Phys., {\bf5}, 110–114, (2009).\\
$^9$ B. Zhao, M. Müller, K. Hammerer, and P. Zoller. Efficient quantum repeater based on deterministic Rydberg gates[J]. Phys. Rev. A, {\bf81}, 052329, (2010).\\
$^10$ A. D. Bounds, N. C. Jackson, R. K. Hanley, R. Faoro, E. M. Bridge,
P. Huillery, and M. P. A. Jones. Rydberg-dressed magneto-optical trap[J]. Phys. Rev. Lett., {\bf120}, 183401, (2018).\\
$^11$ J. D. Carter, O. Cherry, and J. D. D. Martin. Electric-field sensing near the surface microstructure of an atom chip using cold Rydberg atoms[J]. Phys. Rev. A, {\bf86}, 053401, (2012).\\
$^12$ L. A. Jones, J. D. Carter, and J. D. D. Martin. Rydberg atoms with a reduced sensitivity to dc and low-frequency electric fields[J]. Phys. Rev. A, {\bf87}, 71–74, (2013).\\
$^13$ J. D. Bai, S. Liu, J. Y. Wang, J. He, and J. M. Wang. Single-photon Rydberg excitation and trap-loss spectroscopy of cold cesium atoms in a magneto-optical trap by using of a 319-nm ultraviolet laser system[J]. IEEE J. Sel. Top. Quant. Electr., {\bf26}, 1600106, (2020).\\
$^14$ J. Junker, P. Oppermann, and B. Willke. Shot-noise-limited laser power stabilization for the aei 10 m prototype interferometer[J]. Opt. Lett., {\bf42}, 755, (2017).\\
$^15$ F. Seifert, P. Kwee, M. Heurs, B. Willke, and K. Danzmann. Laser power stabilization for second-generation gravitational wave detectors[J]. Opt. Lett., {\bf31}, 2000–2002, (2006).\\
$^16$ J. J. Du, W. F. Li, G. Li, J. M. Wang, and T. C. Zhang. Intensity noise suppression of light field by optoelectronic feedback[J]. Optik, {\bf124}, 3443–3445, (2013).\\ 
$^17$ R. Sun, X. Wang, K. Zhang, J. He, and J. M. Wang. Influence of laser intensity fluctuation on single-cesium atom trapping lifetime in a 1064-nm microscopic optical tweezer[J]. Appl. Sci., {\bf10}, 659, (2020).\\
$^18$ P. Kwee, B. Willke, and K. Danzmann. Shot-noise-limited laser power stabilization with a high-power photodiode array[J]. Opt. Lett., {\bf34}, 2912–2914, (2009).\\
$^19$ Y. Wang, K. Wang, E. F. Fenton, Y. W. Lin, K.-K. Ni, and J. D. Hood. Reduction of laser intensity noise over 1 mhz band for single atom trapping[J]. Opt. Express, {\bf28}, 31209, (2020).\\
$^20$ S. Inoue and Y. Yamamoto. Longitudinal-mode-partition noise in a semiconductor-laser-based interferometer[J]. Opt. Lett., {\bf22}, 328–330, (1997).\\
$^21$ D. Yelin, B. E. Bouma, and G. J. Tearney. Generating an adjustable three-dimensional dark focus[J]. Opt. Lett., {\bf29}, 661–663, (2004).\\ 
$^22$ L. Isenhower, W. Williams, A. Dally, and M. Saffman. Atom trapping in an interferometrically generated bottle beam trap[J]. Opt. Lett., {\bf34}, 1159–1161, (2009).\\
$^23$ Y. H. Gao, Y. J. Li, J. X. Feng, and K. S. Zhang. Stable continuous-wave single-frequency intracavity frequency-doubled laser with intensity noise suppressed in audio frequency region[J]. Chinese Phys. B, {\bf28}, 094204, (2019).\\
$^24$ T. A. Savard, K. M. O’hara, and J. E. Thomas. Laser-noise-induced heating in far-off resonance optical traps[J]. Phys. Rev. A, {\bf56}, R1095, (1997).\\
$^25$ J. D. Bai, S. Liu, J. He, and J. M. Wang. Towards implementation of a magic optical-dipole trap for confining ground-state and Rydberg-state cesium cold atoms[J]. J. Phys. B: At. Mol. Opt. Phys., {\bf53}, 155302, (2020).\\
$^26$ J. D. Bai, X. Wang, X. K. Hou, W. Y. Liu, and J. M. Wang. Angle-Dependent Magic Optical Trap for the $6S_ {1/2}-n P_{3/2}$ Rydberg Transition of Cesium Atoms[J]. Photonics, {\bf9}, 303, (2022).\\ 
$^27$ Y. Y. Wang, X. J. Zhu, J. Liu, Y. B. Ma, Z. H. Zhu, J. W. Cao, Z. H. Du,
X. G. Wang, J. Qian, C. Yin, Z. Y. Liu, D. Blair, L. Ju, and C. N. Zhao. The laser interferometer gravitational wave detector[J]. Progress in Astronomy, {\bf32}, 348, (2014).(In Chinese)\\
\end{document}